\documentclass[12pt]{article}

\usepackage{a4wide}
\usepackage{amssymb}
\usepackage{bm}
\usepackage{latexsym}
\usepackage{dcolumn}
\usepackage{amsfonts,amssymb}
\usepackage{graphicx,epsfig}
\usepackage{psfrag}
\usepackage{amsmath,amssymb}

\def\be {\begin{equation}}
\def\ee {\end{equation}}
\def\beq{\begin{equation}}
\def\eeq{\end{equation}}
\def\bea {\begin{eqnarray}}
\def\eea {\end{eqnarray}}
\def\br{\begin{eqnarray}}
\def\er{\end{eqnarray}}
\def\bc {\begin{center}}
\def\ec {\end{center}}
\def\bfg {\begin{figure}}
\def\efg {\end{figure}}
\def\bi {\begin{itemize}}
\def\ei {\end{itemize}}
\def\benu{\begin{enumerate}}
\def\eenu{\end{enumerate}}
\newcommand{\bdm}{\begin{displaymath}}
\newcommand{\edm}{\end{displaymath}}

\def\l{\left}
\def\r{\right}

\def\laq{\hbox{~}\raise 0.4ex\hbox{$<$}\kern -0.8em\lower 0.62ex\hbox{$\sim$}\hbox{~}}
\def\gaq{\hbox{~}\raise 0.4ex\hbox{$>$}\kern -0.7em\lower 0.62ex\hbox{$\sim$}\hbox{~}}

\def\MPl{M_{_{\rm Pl}}^2}

\newsavebox{\blambox}\savebox{\blambox}[0.6em]{$\lambda\!\!\!$\raisebox{0.5em}
{$\neg$}}\newcommand{\blambda}{\usebox{\blambox}}
\newsavebox{\bFox}\savebox{\bFox}[0.6em]{$F\!\!\!\!$\raisebox{0.5em}
{$\neg$}}\newcommand{\bF}{\usebox{\bFox}}
\newsavebox{\bxibox}\savebox{\bxibox}[0.6em]{$\xi\!\!\!$\raisebox{.5em}
{$\neg$}}

\newcommand{\bphi}{\overline{\varphi}}
\newcommand{\dphi}{\delta \varphi}
\newcommand{\Hm}{\mathcal{H}}

\newcommand{\Fpt}{{\mathcal{F}}}
\newcommand{\Fpteps}{{\mathcal{F}}_{\varepsilon}}

\begin{document}
\begin{titlepage}
\pagestyle{empty}
\baselineskip=21pt
\vspace{2cm}
\rightline{\tt arXiv:~0905.2573}
\vspace{1cm}
\begin{center}
{\bf {\Large 
What-if inflaton is a spinor condensate?
}}
\end{center}
\begin{center}
\vskip 0.2in
{\bf S. Shankaranarayanan}
\vskip 0.1in
{\it 
Institute of Cosmology and Gravitation, 
University of Portsmouth, \\
Mercantile House, Portsmouth P01 2EG, U.K.}\\
{\tt Email: shanki.subramaniam@port.ac.uk}\\
\end{center}

\vspace*{0.5cm}

\begin{abstract}
In the usual cosmological inflationary scenarios, the scalar field --
inflaton --- is usually assumed to be an elementary field. In this
essay, we ask: What are the observational signatures, if
the scalar field is a spinor condensate? and Is there a way to
distinguish between the canonical scalar field and the spinor
condensate driven models? In the homogeneous and isotropic background,
we show that --- although the dark-spinor (Elko) condensate leads to the
identical acceleration equation as that of the canonical scalar field
driven inflation --- the dynamics of the two models are different. In the
slow-roll limit, we show that the model predicts a running of scalar
spectral index consistent with the WMAP data. We show that the
consistency relations between the spinor condensate and canonical
scalar field driven model are different which can be tested using the
future CMB and gravitational wave missions.
\end{abstract}

\vspace*{2.0cm}

\begin{center}
{\bf Essay received an honorable mention from the \\ Gravity Research 
Foundation essay competition in 2009} 
\end{center}

\vfill\vfill

\end{titlepage}

\baselineskip=18pt

Inflation is currently considered to be the best paradigm to describe
the early universe. One of the main reasons is that inflation is
deeply rooted in the basic principles of general relativity and
(quantum) field theory, which are well-tested theories. All the forms
of energy gravitate in general relativity that one of them, the
pressure -- which can be negative in field theory --- is able to cause
the acceleration in the expansion of the universe.  In addition, the
inflationary epoch magnifies the tiny fluctuations in the quantum
fields into classical perturbations that provides a natural
explanation for the origin of the large scale structures and the
associated temperature anisotropies in the Cosmic Microwave Background
(CMB) radiation \cite{Komatsu2008}.

Although it is not entirely justified, it is usually assumed that the
field which drives inflation, {\it the inflaton}, is an elementary
scalar field with no internal structure.  Based on this assumption and
that the field evolves slowly, inflation predicts that: (i) the power
spectrum of the density fluctuations is almost scale invariant and has
no running. (ii) that existence of background primordial gravitational
waves and (ii) ratio of tensor to scalar power spectra is $r = 16
\varepsilon_{\rm can}$ where $\varepsilon_{\rm can}$ is slow-roll
parameter. It has been argued that such a relation, if observationally
verified, would offer strong support for the idea of inflation.

In this essay, we critically analyze this claim by considering a model
in which the inflation is not an elementary field. More precisely, we
ask the following question: If the inflaton is not an elementary
field, how robust are these predictions? It is long known that the
role played by inflaton can also be played by the curvature scalar $R$
\cite{Starobinsky1980}, or logarithm of the radius of compactified
space \cite{Shafi:1986vv} or vector meson condensate \cite{Ford1989}
or a fermionic condensate. Although there has been intense activity in
several of these cases recently \cite{Ford1989}, the possibility of
fermionic condensate has not been discussed much in the literature
\cite{Boehmer2007i,Boehmer2008}.

In this essay, we show that a spinor condensate is a viable
alternative model for scalar driven inflation. In the homogeneous and
isotropic Friedman-Robertson-Walker (FRW) space-time, the spinor
condensate has the identical acceleration equation while the Friedman
equation is modified. In the perturbed FRW background, we show that
(i) scalar and tensor spectra are nearly scale invariant and lead to
running of indices, (ii) The tensor-to-scalar ratio is different
compared to the scalar field driven model.

Before we go in to the details let us ask: Can a scalar condensate
form in the early universe? Free spinors form scalar condensate when
the interacting term (usually 4-Fermi interaction) dominates and
drives the system to a new non-perturbative ground state. The
transition from the free fermions to a scalar condensate occurs below
the critical temperature $T_c$ given by the relation \cite{TC}:
\begin{equation}
T_c \sim \frac{\hbar^2}{k_B}  \frac{\rho^{2/3}}{m^{5/3}} 
\end{equation}
where $\rho$ is energy density and $m$ is mass. At the start of the
inflation, $\rho \sim 10^{78} {\rm GeV}^4$ and $m \sim 10^{16} {\rm
GeV}$, then the critical temperature is $T_c \sim 10^{18} {\rm GeV}$
which is greater than the energy scale of inflation. In the rest of
this essay, we will assume that this spinor condensate dominates
during the early universe and it drives inflation.

But, what kind of spinors can form such a condensate? We know that
electrical conductivity in the universe during inflation was
negligibly small as there were very few charged particles. So, the
spinors we need to consider should not carry charge which means they
are not Dirac spinors. (Dirac spinors, like charged leptons and
quarks, satisfy the electric charge conservation
\cite{weinberg}). Majorana spinors represent particles which do not
carry charge. Although these spinors were known for a long time,
however, the field theory for such a spinor was constructed only
recently \footnote{In Ref. \cite{weinberg}, Weinberg shows explicitly
that conservation of parity as a physical principle to get the Dirac
spinors and constructs its field theory. While, he shows that the
eigen spinors of charge conjugation operators are Majorana spinors, he
does not construct a field theory for the same.}.

Recently, the field theory for the eigen spinors of charge conjugation
operator (Majorana spinors) were constructed by Ahluwalia-Khalilova
and Grumiller \cite{Ahluwalia2005,Ahluwalia:2008xi} and referred them
as Elkos. They showed that these new spinors possess special
properties under discrete symmetries like Charge ${\cal C}$, parity
${\cal P}$ and Time ${\cal T}$ operators. More precisely, they showed
that ${\cal P}^2 = -1, [{\cal C}, {\cal P}] = 0, ({\cal C}{\cal P}
{\cal T})^2 = -1$. The mass dimensions of these spinors, unlike Dirac
spinors, is 1. Under the Wigner classification of field theories, Elko
fall in the category of non-standard Wigner class of spinors
cite{Weinberg}. It is interesting to note that, although Elko satisfy
$({\cal C}{\cal P} {\cal T})^2 = -1$ (i.e. CPT is an anti-unitary
operator), there exist a local quantum field theory
\cite{Streater-Wighman}. 

Since CPT is an anti-unitary for Elko while CPT is unitary for the
standard model particles, it severely restricts the interactions
between Elko and standard matter particle. In other words, the
interactions between Elko and the standard matter particles will
always need Elko and its conjugate. The mass dimension also severely
restrict the kind of interactions they can have with standard model
particles. The spinors have mass dimension one and therefore the only
power counting renormalizable interactions of this field with standard
matter take place through the Higgs doublet or with gravity
\cite{Ahluwalia2005}. Hence, these spinors are good dark matter
candidates \cite{Ahluwalia2005} and are also refereed as dark spinors.

Let us now go to the details and see how the Elko condensate can lead
to inflation and generate primordial perturbations consistent with the
CMB observations but different from the standard scalar field driven
inflation. In order to do this, we need to know the Lagrangian and the
form of Elko. Since the mass-dimension of Elko is unity, its
Lagrangian is like that of a scalar field:
\beq
\label{eq:ElkoLag}
\mathfrak{L}_{elko}= 
\frac{1}{2} 
\l[\frac{1}{2} g^{\mu \nu}(\mathfrak{D}_\mu \blambda \mathfrak{D}_\nu \lambda 
+ \mathfrak{D}_\nu \blambda \mathfrak{D}_\mu \lambda) \r] 
- V(\blambda \lambda) \, .
\eeq
where $\mathfrak{D}_\nu$ is the covariant derivative given by
\beq
\mathfrak{D}_\mu \lambda = 
(\overrightarrow{\partial}_\mu + \Omega_\mu)\,\lambda(x)~;~
\mathfrak{D}_\mu\blambda = 
\blambda(x) \, (\overleftarrow{\partial}_\mu - \Omega_\mu) \, ,
\eeq
$\Omega_{\mu}$ is the tangent space connection and $\gamma$'s are the
Dirac matrices in the Weyl representation [For more details, see
Refs. \cite{Boehmer2007i,ours}.]  Elko $\lambda(x)$ and its dual
$\blambda(x)$ have the following form:
\begin{equation}
\label{eq:elkosform}
\lambda(x) =\left(\begin{array}{c} 
              \pm \sigma_2 {\phi^{(1)}}^* \\ 
              \phi^{(1)} 
              \end{array} \right)
\qquad 
\blambda(x) = i \left({\phi^{(2)}}^{\dagger} 
                   \pm {\phi^{(2)}}^{\dagger} \sigma_2 
             \right)
\end{equation}
We now have all the armory to study the evolution of perturbations of
the Elko. To linear order in fluctuations (and neglecting vector
modes), the line element --- in the longitudinal gauge --- for a
spatially flat FRW background can be written as
\cite{Kodama-Sasa:1984}:
\beq
\label{eq:LinearFRW} 
{\rm d}s^2 = 
a^2 \, \{(1 + 2 \Phi ){\rm d}\eta ^2  
- [(1- 2 \Psi )\delta _{ij} + h_{ij}]{\rm d}x^i{\rm d}x^j\}\, , 
\eeq
where $\Phi, \Psi$ are the Bardeen potentials and represent the scalar
sector, and the traceless and transverse tensor $h_{ij}$
($h_i{}^i=h_{ij}{}^{,j}=0$), represents the tensor sector, i.e. the
gravitational waves. $\eta = \int [dt/a(t)]$ is the conformal time,
the Hubble rate is $H \equiv \dot{a}/a = {\cal H} a$, ${\cal H}
\equiv {a}'/a$ and a prime refers to derivative with respect to
$\eta$.

In the case of homogeneous-isotropic FRW background, the Einstein
equations demand that $\lambda$ and $\blambda(x)$ should depend on
single scalar function cite{Boehmer,our} i. e
\beq
\lambda(x)  = \overline{\varphi}(\eta) \lambda_0;\quad
\blambda(x) = \overline{\varphi}(\eta) \lambda_1
\eeq
where $\lambda_0, \lambda_1$ are constant column and row vectors,
respectively. The acceleration and Friedman equations are then given
by:
 \br 
\label{eq:Eacceleration}
& & {\cal H}' = \frac{1}{3 \MPl} 
\l[a^2(\eta) V(\overline{\varphi}) - {\bphi'}^2(\eta)\r] \\
\label{eq:EFriedmann}
& &  \mathcal{H}^2 = \frac{1}{3 \MPl} 
\l[\frac{{\bphi'}^2/2 
      + a^2(\eta) V(\overline{\varphi})}{1 + \Fpt}\r] 
\qquad  \quad \Fpt = \frac{\bphi^2}{8 \MPl} 
\er
The following points are worth noting regarding the above results: \\
(i) The background Elko (and its dual) depend on a single scalar
function $(\overline{\varphi})$ satisfying $ \overline{\blambda}
\overline{\lambda} = \overline{\varphi}^2(\eta)$. This can be
interpreted as an Elko-pair forming a scalar
condensate --- spinflaton. \\
(ii) The acceleration equation for the spinflaton
(\ref{eq:Eacceleration}) is identical to canonical scalar field
inflation. However, the Friedmann (\ref{eq:EFriedmann}) equations have
non-trivial corrections due to Elko. The Elko modification to the
canonical inflaton equations are determined by $\Fpt$. \\
(iii) In order to see the nature of the inflationary solutions, it is
illustrative to look at power-law inflation i. e. $a(t) \propto t^n
~~(n > 1)$. In this case, we have
\beq
V(\varphi) = 3 \MPl \overline{\varphi}^{\pm 1/n} + c_0 \overline{\varphi}^{2 (n - 2)/n} 
\qquad \quad \overline{\varphi} \propto t^{\pm n/2}  
\eeq
The form of the potential and the evolution of the field is completely
different compared to the canonical scalar field. {\it In other words,
the dynamics of spinflaton is different to the inflationary scenario
assuming that the scalar field in elementary.} Note that the above
potential satisfies the slow-roll parameters defined in
Ref. \cite{ours}.

Having showed that the spinflaton dynamics is different compared to
the canonical scalar field, our next step is to look at the evolution
of metric perturbations (\ref{eq:LinearFRW}) in spinflation. At the
linear level, scalar and tensor perturbations decouple and can be
treated separately. As in standard inflation, the tensor perturbations
do not couple to the energy density and pressure inhomogeneities. They
are free gravitational waves and satisfy:
\beq
\label{eq:TensorPertu}
\mu_{_T}'' + \l(k^2 - \frac{a''(\eta)}{a(\eta)}\r)\mu_{_T} = 0 \, .
\eeq

The main hurdle in obtaining perturbation equation for the scalar
perturbations is the perturbed stress-tensor for the Elko. Unlike the
scalar field, which has only one degree of freedom, Elko, in general,
has four complex functions; not all of these are independent and are
related by constraints. The most general perturbed Elko can lead to
the scalar and vector perturbations. Besides, it can also lead to a
non-vanishing anisotropic stress. For simplicity, we will assume that
the anisotropic stress is identically zero.

Perturbation of a spinor has long been studied in spherically
symmetric Skyrmion model using the hedgehog ansatz
\cite{Chodos1975}. We use the similar procedure for the perturbed Elko
and its dual in the case of perturbation theory i. e.
\beq
\label{eq:peransatz}
\delta \lambda(x^\mu)= i \, F[\gamma] \, \overline{\lambda}(\eta) 
\qquad \delta \blambda(x^\mu) = -i \, \overline{\blambda}(\eta) \, 
\l[\gamma^0 \, \bF[\gamma]^\dagger \, \gamma^0 \r]   
\eeq
where 
$F[\gamma], \bF[\gamma]$ are chosen such that it leads to a consistent 
perturbation equations cite{ours}. As in the canonical scalar field, 
the perturbation equations can be combined in to a single equation 
in-terms of modified Mukhanov-Sasaki variable, i. e.,
\beq
\label{eq:MSequation}
Q'' - \l[\nabla^2  + \frac{z''}{z} 
        - \ln[1-\Fpteps]'' 
        + \frac{7 \Hm' \Fpteps^{\frac{1}{2}} }{2}
        + \frac{\Hm \varepsilon' 
       \Fpteps^{\frac{1}{2}}}{\varepsilon}\r] Q \simeq 0 
\eeq 
where $\varepsilon$ is the slow-roll parameter.
\beq
Q =  a \, \dphi + z \, \Psi;~~ z = \left[1- \Fpteps \right] (a \bphi')/\Hm~;~~
\Fpteps = \frac{\cal F}{{\cal F} + \varepsilon}
\eeq
The above equation is the first key result of this essay. We would
like to stress the following points: \\
(i) We have ignored the contributions of higher-order slow-roll
functions like $\varepsilon^2 \Fpteps^{1/2}, \Fpteps$. In the leading
order slow-roll approximation, their contribution to the scalar power
spectrum is tiny. \\
(ii) The Elko modification to the canonical MS equation is determined
by $\Fpt$. \\
(iii) The sound speed of perturbations for the Elko is identical to
that of the canonical scalar field.\\
(iv) In the slow-roll limit, the non-adiabatic part of the
perturbations vanishes on super-Hubble scales i.e., the entropy
perturbation $\propto \nabla^2 \Psi$.

Invoking the slow-roll conditions ($\varepsilon, |\delta| \ll 1$) in
the scalar (\ref{eq:MSequation}) and tensor perturbation equations
(\ref{eq:TensorPertu}) --- and following the standard quantization
procedure by assuming Bunch-Davies vacuum --- the scalar and tensor
power spectra are given by:
{\small
\br
\label{eq:SPS}
{\cal P}_S(k) & =& \l(\frac{H^2}{8 \MPl \pi^2}\r)
\left(\frac{\varepsilon + \Fpt}{\varepsilon^2}\right) 
\l[1- 2(c_0+1)\varepsilon_{_{\rm can}} \r] \\
\label{eq:TPS}
{\cal P}_T(k) &=& \l(\frac{2 H^2}{\MPl \pi^2}\r) 
\l[1- 2(c_0+1) \varepsilon_{_{\rm can}} + 2 \varepsilon_{_{\rm can}} x\r]
\er
}
\hspace*{-7pt} 
where $c_0=\gamma_{_{\rm Euler}}+ \ln2- 2$ is a constant, $x=\ln(k^*/k)$
and $k^*$ is the pivot scale.

Eqs. (\ref{eq:SPS}, \ref{eq:TPS}) allow us to draw important
conclusions which are the second key result of this essay. Firstly,
the scalar and tensor power spectra of the spinflaton, in slow-roll,
are nearly scale-invariant \cite{ours}. Secondly, in the leading order
of $\epsilon_{_{\rm can}}$, the running of scalar spectral index is
non-zero and given by
{\small
\begin{subequations}
\br
\frac{d n_S}{d \ln{k}} &=&
-\frac{\varepsilon_{_{\rm can}}}{2}- 4 \varepsilon_{_{\rm can}}
\Fpteps^{1/2} + \frac{\varepsilon_{_{\rm can}}}{2}\frac{\Fpt}{1+\Fpt} 
\er
\end{subequations}
}
\hspace*{-7pt} It is interesting to note that the running of scalar spectral 
index ($-0.09 < d n_S/d (\ln k) < 0.019$) is consistent with the
WMAP-5 year results \cite{Komatsu2008}. For instance, using the WMAP
value of $\varepsilon_{_{\rm can}} = 0.038$ \cite{Komatsu2008} and
assuming that $\Fpt$ is tiny, we get $d n_S/d (\ln k) \sim
-0.019$. This is the one of the main predictions of spinflation.

Thirdly, the tensor-to-scalar ratio $r$ is no longer equal to $16
\varepsilon_{_{\rm can}}$ and is given by:
\beq
\label{eq:r}
r \simeq 16 \, \varepsilon_{_{\rm can}} \, \l[1- 2 \Fpteps\r] \, .
\eeq
Physically, this suggests that the gravitational wave contribution
during slow-roll spinflation is smaller than for canonical
inflation. Lastly, as for the canonical scalar field, the scalar and
tensor perturbations during spinflation originate from the scalar
condensate and they are not independent. Hence, consistency relations
link them together. The one which is observationally useful is the
relation between $n_{_T}$ and $r$:
{\small
\beq
n_{_T}=\frac{r}{8} (1+\Fpteps) \l[ 1  
               + \varepsilon_{_{\rm can}}\l[\frac{11}{6} c + \Fpteps-\Fpt \r] 
               - 2\delta_{_{\rm can}} \, c \r] \, .
\eeq
}
\indent To conclude, we have shown that the spinor condensate 
in the early universe is a viable model of cosmological inflation.  It
leads to the identical acceleration equation as that of canonical
scalar field driven inflation. We have used the Hedgehog ansatz to
obtain the scalar perturbation and, in the slow-roll limit, we have
shown that scalar and tensor perturbations are nearly scale
invariant. The model predicts a running of scalar spectral index
consistent with WMAP-5 year data. The consistency relation between the
scalar and tensor spectra are non-trivial and have different feature
compared to the models where the scalar fields are considered
elementary.

The author wishes to thank Damien Gredat, Christian Boehmer and Roy
Maartens for discussions. The work is supported by the Marie Curie
Incoming International Grant IIF-2006-039205.



\end{document}